\documentclass[aip, apl, superscriptaddress, reprint, floatfix]{revtex4-1}
\usepackage[utf8]{inputenc}
\usepackage[english]{babel}
\usepackage{graphicx}
\begin{document}
\title{Atomic-scale identification of novel planar defect phases in heteroepitaxial YBa$_2$Cu$_3$O$_{7-\delta}$ thin films}

\author{Nicolas Gauquelin}
\email[]{Nicolas.Gauquelin@uantwerpen.be}
\thanks{Current Address: EMAT, Department of Physics, University of Antwerp, Groenenborgerlaan 171, 2020
	Antwerp, Belgium}
\affiliation{Canadian Centre For Electron Microscopy, McMaster University, 1280 Main Street West, Hamilton, Ontario, L8S 4M1, Canada}

\author{Hao Zhang}
\affiliation{Department of Physics, University of Toronto, 60 St. George Street, ON M5S1A7, Toronto, Canada}

\author{Guozhen Zhu}
\affiliation{Canadian Centre For Electron Microscopy, McMaster University, 1280 Main Street West, Hamilton, Ontario, L8S 4M1, Canada}

\author{John Y.T. Wei}
\affiliation{Department of Physics, University of Toronto, 60 St. George Street, ON M5S1A7, Toronto, Canada}
\affiliation{Canadian Institute for Advanced Research, Toronto, M5G1Z8, Canada}

\author{Gianluigi A. Botton}
\email[permanent corresponding author:]{gbotton@mcmaster.ca}
\affiliation{Canadian Centre For Electron Microscopy, McMaster University, 1280 Main Street West, Hamilton, Ontario, L8S 4M1, Canada}
\begin{abstract}
  We have discovered two novel types of planar defects that appear in heteroepitaxial YBa$_2$Cu$_3$O$_{7-\delta}$ (YBCO123) thin films, grown by pulsed-laser deposition (PLD) either with or without a La$_{2/3}$Ca$_{1/3}$MnO$_3$ (LCMO) overlayer, using the combination of high-angle annular dark-field scanning transmission electron microscopy (HAADF-STEM) imaging and electron energy loss spectroscopy (EELS) mapping for unambiguous identification. These planar lattice defects are based on the intergrowth of either a BaO plane between two CuO chains or multiple Y-O layers between two CuO$_2$ planes, resulting in non-stoichiometric layer sequences that could directly impact the high-$T_c$ superconductivity.
\end{abstract}

\maketitle

Recently, a great deal of interest has been focused on thin-film heterostructures comprising YBa$_2$Cu$_3$O$_{7-\delta}$ (YBCO123) and other complex oxides, because of novel interfacial interactions\cite{Zhang2009,Varela2003,Habermeier2001} that could affect the high-critical temperature ($T_c$) superconductivity.\cite{Bacca2004,Chakhalian2007}YBCO123 in heteroepitaxial form is known to contain a variety of crystalline defects, as its layered perovskite structure is sensitive to lattice strains induced by the epitaxial mismatch.\cite{Ramesh1992,Zhang2013a,Fendorf1990,Foltyn2007,Grigis1999,Domenges1991} Since such defects may also affect the microscopic pairing mechanism, it is important to determine their lattice structures at the atomic scale.  There is solid experimental evidence to suggest that the interfacial lattice mismatch in heteroepitaxial thin films of complex oxides can, through induced lattice strains, affect the microstructure in the film.

In an earlier study of YBCO123 thin films heterostructured with La$_{2/3}$Ca$_{1/3}$MnO$_3$  (LCMO) overlayers, we observed nanoscale domains of YBCO124 and YBCO247 containing CuO-chain intergrowths, which were attributed to the heteroepitaxial strain and also shown to cause attenuation of $T_c$.\cite{Zhang2013a} In the present study, we focus on defect phases associated with Ba-O and Y-O intergrowths in heteroepitaxial YBCO123 thin films both with and without LCMO overlayers.  Two novel types of planar defects are unequivocally observed, both deviating from the normal stacking sequence (CuO-BaO-CuO$_2$-Y-CuO$_2$-BaO) of the YBCO123 lattice.  The first defect type (henceforth called type \textbf{D1}) consists of a BaO layer sandwiched between two CuO-chain layers.  The second defect type consists of the addition of either one (type \textbf{D2}) or two (type \textbf{D2'}) extra Y-O layers between the CuO$_2$ planes.

The samples used in this study include single-layer YBCO123 and bi-layer LCMO/YBCO123 thin films.  The layer thickness of YBCO123 was either 25 nm or 50 nm, while the LCMO overlayer thickness was 25 nm. The films were epitaxially grown on (001)-oriented (La,Sr)(Al,Ta)O$_3$ (LSAT) substrates using pulsed laser-ablated deposition (PLD). The films were annealed {\it in-situ} by slowly cooling from $800\,^{\circ}{\rm C}$ to $300\,^{\circ}{\mathrm C}$ in 1 atm of O$_2$ at a rate of $11\,^{\circ}{\rm C}$/min. Cross-section of the resulting samples were then prepared by FIB and thinned with an Ar beam to 40 - 45 nm in a Fischione Nanomill 1040 instrument operated at 900eV. Imaging by scanning transmission electron microscopy (STEM) and spectroscopy of electron energy loss (EELS) were carried out with an FEI Titan$^{3}$  TEM, equipped with a CEOS-designed hexapole-based aberration corrector for both the probe-forming lens and imaging lens. The instrument is fitted with a high brightness electron source, in order to achieve sub-angstrom resolution \cite{Botton2010}. Measurements were performed at 200 keV in order to reduce knock-on damage. EELS spectra were recorded on a high resolution Gatan GIF Tridiem spectrometer. The convergence semi-angle for STEM was 20 mrad, the inner acceptance semi-angle for High Angle Annular Dark Field (HAADF) imaging was 50 mrad, the collection semi-angle for EELS was 110 mrad.

\begin{figure}[htb]
   \includegraphics[width=0.8\columnwidth]{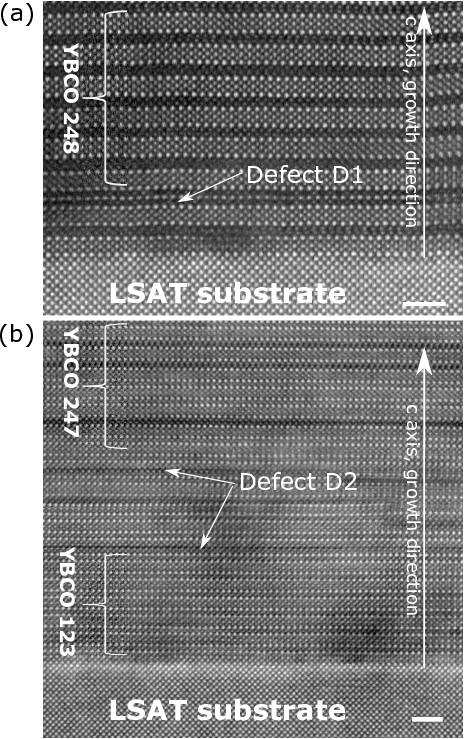}
   \caption{HAADF STEM images along the [001] crystallographic direction of single layer YBCO123 films. Scale bars are 2nm.  Local domains of YBCO248 and YBCO247, resulting from intergrowth of CuO chains, are present in the matrix of YBCO123. The defects highlighted by arrows are of two type: a) defect \textbf{D1} consists of the insertion of a BaO plane in between two CuO$_{1-\delta}$ chains and b) defect \textbf{D2} is related to the insertion of one YO$_2$ layer between the CuO$_2$ planes.}
    \label{Fig:haadf}
\end{figure}

Fig. ~\ref{Fig:haadf} shows $Z$-contrast STEM images (in which the intensity is proportional to $Z^{1.7}$, and $Z$ is the atomic number) taken using the HAADF technique on single-layer YBCO123 films. As was seen in our previous study, local domains of YBCO248 and YBCO247, resulting from intergrowth of CuO chains, are present in the matrix of YBCO123. In the case of orthorhombic YBCO123 grown on and between cubic perovskites, Cu-O stacking faults have previously been observed \cite{Fendorf1990} and are associated with the formation of defect phases such as YBCO124, YBCO247, YBCO125 and YBCO249 (for which the first number corresponds to the number of Y in a unit cell and the subsequent second and third number Ba and Cu, respectively), depending on the number of CuO chains per unit cell.\cite{Ramesh1992,Grigis1999} Also visible in Fig. ~\ref{Fig:haadf}(a) are type \textbf{D1} defects, which is related to an excess BaO layer between two CuO$_x$ chains.  In previous studies \cite{Ramesh1992,Grigis1999}, these defects were interpreted mainly as YO-CuO or as BaO/YO interconversion from HRTEM image analysis; however, Domenges {\it et al}.\cite{Domenges1991} has pointed out the unfavourable nature of the YO-CuO structure, which would distort the environment of the YO$_8$ cage. More recently, Gazquez {\it et al}.\cite{Gazquez2012b} have interpreted this defect type in terms of Ba-CuOx intergrowth.  In our present study, atomic-scale HAADF imaging was used in conjunction with elemental EELS mapping to elucidate the exact lattice structure and atomic composition of the \textbf{D1} defect structure, as we present below.

\begin{figure}[htb]
   \includegraphics[width=0.8\columnwidth]{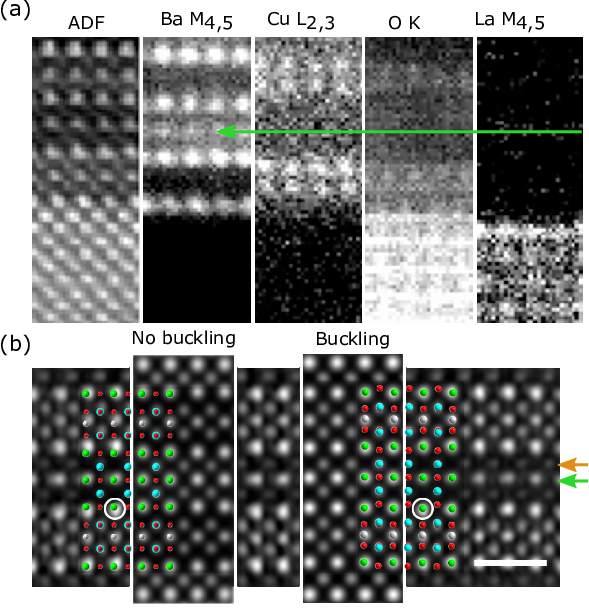}
   \caption{This figure will appear in color online. a) Atomic-resolved EELS elemental maps of Ba, Cu, O and La at the interface between LSAT substrate and YBCO of a LCMO(25 nm)/YBCO(25 nm) double layer film on LSAT substrate.  The same type of defect seen in Fig. ~\ref{Fig:haadf}(a) is visible here, and can be identified as a BaO-CuO$_{1-\delta}$ intergrowth. The long green arrow indicates the presence of an extra Ba-containing layer. b) Noise filtered HAADF image showing the \textbf{D1} defect; Green and red arrows indicate BaO and CuO$_{1-\delta}$ layers respectively. Scale bars are 1 nm. Overlay in b show the structural models with Ba in green, O in red, Cu in turquoise and Y in grey and the simulated images based on the YBa$_2$Cu$_3$O$_6$ lattice, without buckling (left overlay) and with buckling (right overlay).}
    \label{Fig:Badefect}
\end{figure}

\begin{figure}[htb]
	\includegraphics[width=1\columnwidth]{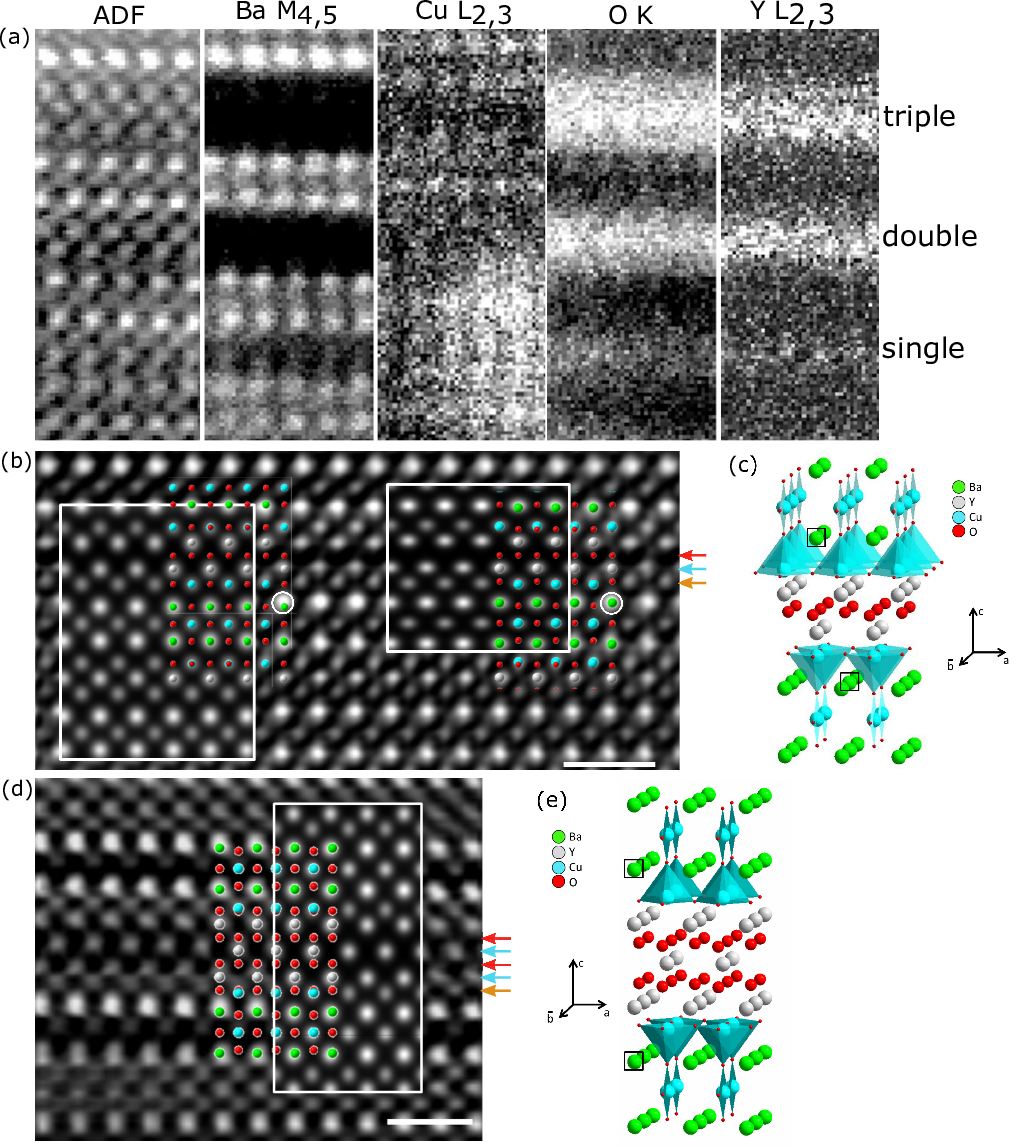}
	\caption{This figure will appear in color online. a) Atomic-resolved EELS elemental maps of Ba, Cu, O and Y in a region of a YBCO single-layer film (25 nm) on LSAT substrate where the \textbf{D2} and $\textbf{D2’}$ defects structures can be identified as Y-O$_2$ and Y-O$_2$-Y-O$_2$ intergrowth respectively from the Y and O EELS mapping. b) HAADF image of a \textbf{D2} and a $\textbf{D2’}$ defect lattice; Blue and red arrows indicate Y and $O_2$ layers from the fluorite layer respectively, the crystal structure model determined from this is shown in panel c. d) HAADF image of a triple-Y $\textbf{D2’}$ defect, overlayed with a simulated image and structural model based on a buckled YBa$_2$Cu$_3$O$_6$ lattice, the crystal structure model determined from this is shown in panel e. Scale bars are 1 nm. Each overlay in b show the structural models with Ba in green, O in red, Cu in turquoise and Y in grey. The simulated images are based on the YBa$_2$Cu$_3$O$_7$ lattice, both without buckling (left overlay) and with buckling (right overlay).}
	\label{Fig:Ydefect}
\end{figure}

As shown in ~\ref{Fig:Badefect}(a), the \textbf{D1} defect type we observed can be systematically identified in terms of BaO-CuO$_{1-\delta}$ intergrowth using atomic-scale EELS mapping whereby an extra Ba layer is detected in the Ba M$_{4,5}$ map (pointed out with a green arrow in ~\ref{Fig:Badefect}(a)). The \textbf{D2} defect type, according to the elemental maps seen on Figure ~\ref{Fig:Ydefect}(a) (deduced from the Y L$_{2,3}$ edge intensity), contain additional Y-O$_2$ layers between the CuO$_2$ planes (shown by red and blue arrows in ~\ref{Fig:Ydefect}(b)). Care was taken such that the application of weighted principal component analysis (PCA) within the multivariate statistical analysis plugin developed by M. Watanabe \cite{Bosman2006} was only increasing the already good contrast in the elemental maps and not altering it or introducing artefacts. In Fig. ~\ref{Fig:Badefect}(b) we first note that the Ba atoms in the BaO layer are displaced away from the CuO chain\cite{Hott2004,Tsuei2000}. This displacement is similar to the buckling distortion seen in the CuO$_2$ plane of all cuprate superconductors, where the four oxygen atoms from the CuO$_2$ plane are displaced towards the Y spacer layer to form a CuO$_5$ pyramid with the apical oxygen atoms. Although buckling of the CuO$_2$ planes is expected to have a stronger effect on the $T_c$  of YBCO123\cite{Sardar2001,Chmaissem1999b} than buckling of the BaO planes, the latter does significantly increase the Ba-Ba separation about the CuO-chain layer.\cite{Okamura1987,Tsuei2000} This increased Ba-Ba separation (easily noticeable from HAADF-STEM images) effectively increases the Cu-O apical bond length, and could thus inhibit the charge transfer between chain and plane. Taking into account these buckling effects on the YBCO123 lattice\cite{Akimitsu1992,Swinnea1987,Sato1988}$^,$\footnotemark[1], we consider two structural models for the \textbf{D1} defect, which are overlaid as simulated images in Fig. ~\ref{Fig:Badefect}(b) for side-by-side comparison with the experimental HAADF images. HAADF images were simulated using the multislice method carried out with the code developed by Kirkland.\cite{Kirkland1991} A Gaussian broadening was applied to these simulated images to consider, as a first approximation, the finite source size and to improve the accuracy of the comparison with the experimental data.\cite{Nellist1994} The experimental HAADF images were Fourier-filtered to reduce the noise.  For the lattice position of Ba, better agreement with the experimental image is obtained by assuming a similar buckling in the outer BaO layers (highlighted by open circle) as in fully-oxygenated YBCO123 (YBa$_2$Cu$_3$O$_7$), and an absence of distortion in the central BaO plane. The larger spacing between the BaO planes on each side of the defect (9.2 \AA\ for empty chains versus 8.7 \AA\ for full chains) indirectly suggests (in conjunction with the absence of heavy atoms from HAADF imaging) that the CuO chains are oxygen-deficient in this particular area (see Supplementary material).  The simulated image with empty chains surrounding the BaO defect layer is shown as an overlay in Figure ~\ref{Fig:Badefect}(b), and is in excellent agreement with the experimental image. The dimmer intensity of both the O K-edge EELS map shown in Fig. ~\ref{Fig:Badefect}(a) and the HAADF images in Fig.  ~\ref{Fig:haadf}(a) gives further indication that this intercalated BaO plane is surrounded by oxygen-deficient CuO$_{1-\delta}$ chains. Therefore, we can infer the defect structure \textbf{D1} as having a local stoichiometry of YBa$_3$Cu$_4$O$_{9-\delta}$. The difference between these values is several times greater than the precision achievable through STEM imaging (around 10pm) making these assumptions on oxygen stoichiometry trustworthy.

The \textbf{D2} defect type we observed is shown in Fig. ~\ref{Fig:Ydefect}(a) and ~\ref{Fig:Ydefect}(b) and also visible in Fig. 1(b), and to our best knowledge has never been reported before in YBCO compounds.  HAADF images of the ribbon structures formed by the double-layered (\textbf{D2})(Y-O$_2$-Y) and triple-layered (\textbf{D2'}) (Y-O$_2$-Y-O$_2$-Y) intergrowths  are shown in Fig. ~\ref{Fig:Ydefect}(b) and ~\ref{Fig:Ydefect}(d) respectively.  It is worth noting that a very dark band between the Y bright layers may indicate the presence of an additional layer (of a low-$Z$ element) not visible by HAADF imaging. Precise structural determination of this defect type is non-trivial, requiring \textit{a priori} knowledge about ideal interplanar distances for YBCO123\cite{Akimitsu1992,Swinnea1987,Sato1988} and structural comparisons with other cuprate superconductors, as discussed below.\footnote{\label{note:note1} According to structural refinement by x-ray crystallography, the Ba atom is vertically displaced by 0.32 \AA\  (0.499 \AA)  relative to the apical O for fully-oxygenated (under-oxygenated) YBCO123 \cite{Sato1988,Okamura1987}. Thus the Ba-Ba distance about the CuO-chain layer is 4.336 \AA\  or 4.598 \AA\ respectively, which is sizable compared with the value of 3.7 \AA\ for an ideal lattice with no buckling\cite{Akimitsu1992}.}.

In a general manner, high-$T_c$ cuprate superconductors can be represented by the formula unit A$_m$E$_2$Y$_{n-1}$Cu$_n$O$_{2n+m+2+y}$ where A and E denote atoms in the EO/(AO$_x$)$_m$/EO charge reservoir (here E=Ba and A=Cu, stacked with \textit{n} CuO$_2$ superconducting planes and \textit{n}-1 Y spacer layers) and conveniently abbreviated as A-\textit{m}2(\textit{n}-1)\textit{n}.\cite{Hott2004} By this nomenclature, YBCO123 can be represented as Cu-1212. The spacing between the oxygen atoms of the CuO$_2$ plane and Y layers is 1.418 \AA\ in YBa$_2$Cu$_3$O$_7$\cite{Sato1988}, due to the buckling of the CuO$_2$ planes discussed earlier (O shifted towards the Y plane). Similarly in the Pr-based cuprate of the Hg-1222 and Tl-1222 families \cite{Domenges1991,Hervieu1995}, a second Pr plane is intercalated between the CuO$_2$ planes, the distance between the Pr and O ions being 1.4 \AA. This spacing is remarkably identical to the spacing observed between each Pr layer and the central O$_2$ layer in this 1222 structure.\cite{Domenges1991,Hervieu1995} In the HAADF image shown in Fig.  ~\ref{Fig:Ydefect}(b), the spacing between the two Y layers is about twice as large (2.8 \AA) as the spacing between the Y and CuO$_2$ planes, suggesting the presence of a similar O$_2$  plane between the Y layers in the \textbf{D2} defect structure.

Using these structural analogies, we establish a model for the double-layered Y-O$_2$ defect as shown in Fig.  ~\ref{Fig:Ydefect}(c).  A model for the triple-layered Y-O$_2$ defect can be similarly established as shown in Fig.  ~\ref{Fig:Ydefect}(e).  Details of our structural models, corresponding to the \textbf{D2} and \textbf{D2'} defect types, are given respectively in Supplementary Materials. HAADF images were also simulated from the \textbf{D2} and \textbf{D2'} defect lattices plotted as overlays in Fig.  ~\ref{Fig:Ydefect}(b) and  ~\ref{Fig:Ydefect}(d) for comparison with the experimental HAADF images.  We note that, as in the case of the BaO defect discussed above, the Ba-Ba spacing shows better agreement between the simulated and experimental images when buckling of the CuO$_2$ and BaO planes is conserved as in the previous case.  In terms of interplanar spacings, the Ba and Cu positions in the experimental image are better explained by assuming oxygen-rich CuO chains on both sides of this defective BaO-CuO$_2$-Y-O$_2$-Y-CuO$_2$-BaO unit (see Supplementary material).

As for the \textbf{D2} defect shown in Fig.  \ref{Fig:Ydefect}(b), comparison of the Ba-Ba spacing between the experimental image and the simulated image for under-oxygenated YBCO123 (YBa$_2$Cu$_3$O$_6$) suggests that these CuO$_{1-\delta}$ chains are oxygen-deficient in this particular area.  Thus the \textbf{D2} and \textbf{D2'} defects correspond to local stoichiometries of Y$_2$Ba$_2$Cu$_3$O$_{10}$ and Y$_3$Ba$_2$Cu$_3$O$_{12-\delta}$ respectively.  Here it should be noted that O K-edge maps cannot be directly used to determine oxygen content, as the contrast is strongly influenced by channeling of the neighboring columns of heavy atoms; nonetheless a hint on the O content in layers with the same environment (or coordination sphere) can be postulated.  In the oxygen mapping of Fig.  \ref{Fig:Ydefect}(a), we do notice a strong increase in the intensity of the $K$-edge map versus the number of Y layers, confirming the presence of oxygen atoms between the Y planes.

It is remarkable to note that, while neither \textbf{D1}, \textbf{D2} and \textbf{D2'} defect structure appears to modify the buckling of the BaO and CuO$_2$ planes, the zigzag ribbon structure of the double Y-O$_2$ layer in the \textbf{D2} case effectively breaks inversion symmetry of the YBCO123 lattice about the Y plane. This structural peculiarity is physically significant, in that the resulting lack of a center of symmetry along the c-axis could profoundly affect the electron pairing in the CuO$_2$ layers\cite{Bauer2012,Tsuei2000}, either by allowing spin-singlet/triplet mixing or by introducing a Rashba spin-orbit term in the pairing Hamiltonian, especially when strong electron correlations are present.  Prior studies of such non-centrosymmetric superconductors have been limited to a few intermetallic materials with rather low $T_c$, such as CePt$_3$Si, Li$_2$Pd$_3$B and KOs$_2$O$_6$\cite{Bauer2004,Yuan2006,Shibauchi2006}. Our observation of the \textbf{D2} defect structure, and its role in forming a YBCO lattice with broken inversion symmetry, suggests that it is potentially useful as a building block for synthesizing non-centrosymmetric high-$T_c$ superconductors via heteroepitaxial oxide engineering. According to geometrical phase analysis (see supplementary material for details) the \textbf{D2} type defect locally affects the $\langle 00l \rangle$ and $\langle 0k0 \rangle$ family of planes but not the $\langle 0kl \rangle$ planes. These \textbf{D2} defects are linked to some antiphase boundary delimiting an intergrowth region. This boundary might act as pinning center for vortices as discussed previously for BaZrO$_3$ or other type of foreign intergrowths.\cite{Gazquez2012b,Llordes2012,Guzman2013} In contrast, the defect \textbf{D1} does not influence any of the three aforementioned families of planes.

Finally, we discuss generic implications of the defect structures identified above, in relation to the superconducting properties of the high-$T_c$ cuprates. First, it should be noted that since the volume fraction of these defects is rather small, a direct measurement of their influence on the superconductivity of the film is thus not possible.  Nevertheless, by virtue of their novel stacking sequence, the primary effects of these defect structures on fully-oxygenated YBCO123 (structure type 1212) can be inferred from analogies with similar defect structures observed in the bulk form of more complex cuprates.  For the \textbf{D1} defect (structure type 2312), carrier localization within this BaO-CuO$_{1-\delta}$-BaO-CuO$_{1-\delta}$-BaO block may occur as a result of inhibited conduction within the oxygen-depleted chains.  For the \textbf{D2} defect (structure type 1222), insertion of YO$_2$ layers (which follows the fluorite structure\cite{Grigoraviciute2007,MARTIN1989}) between the CuO$_2$ planes increases the spacing between the two CuO$_2$ planes, and could thus directly affect $T_c$ as in the case of Tl- and Hg-based cuprates.\cite{Domenges1989,Hervieu1995} Furthermore, for the \textbf{D2} defect, it is conceivable for excess oxygen within the YO$_2$ layers to more tightly bind together the CuO$_2$ layers and reduce their spacing to the Y plane, thus enhancing the hole concentration within them.\cite{Abrikosov1999,Bianconi2010} Since the high-$T_c$ cuprates that contain these fluorite-type defects are predominantly tetragonal, we can infer that the presence of these defects in YBCO123 tends to weaken its orthorhombicity, thereby reducing the oxygen content and thus the hole concentration in the CuO chains.

In summary, we have performed atomic-scale STEM, via Z-contrast HAADF imaging and elemental EELS mapping, of defect structures in heteroepitaxial YBCO123 thin-film samples grown on LSAT substrates by PLD.  We focused on the observation of two novel types of planar defects associated with Ba-O and Y-O intergrowths, and used atomic-scale data analysis with model simulations to unequivocally identify their lattice structures. One defect type (\textbf{D1}) consists of a BaO layer between two oxygen-deficient CuO$_{{1-\delta}}$ chains corresponding to the formula unit of YBa$_4$Cu$_4$O$_{9-\delta}$; the other defect type (\textbf{D2} and \textbf{D2'}) consists of either a Y-O$_2$-Y or Y-O$_2$-Y-O$_2$-Y fluorite-like block sandwiched between the CuO$_2$ planes, corresponding to Y$_2$Ba$_2$Cu$_4$O$_{10}$ and YBa$_3$Cu$_4$O$_{12-\delta}$ respectively. While the \textbf{D1} defect is expected to localize charge carriers in the CuO chains, the \textbf{D2} and \textbf{D2'} defects are expected to directly impact the high-$T_c$ pairing by increasing the CuO$_2$ bilayer spacing.  Remarkably, the \textbf{D2} defect is shown to break inversion symmetry of the YBCO123 lattice, and could thus serve as a building block for synthesizing non-centrosymmetric high-$T_c$ superconductors. Our STEM study demonstrates that the combined use of atomic-scale HAADF imaging and EELS mapping can effectively identify defect structures in complex materials.

\section*{Acknowledgements}
We are thankful to Julia Huang for FIB TEM sample preparation. This work is supported by NSERC (through Discovery Grants to GAB and JYTW) and CIFAR. The electron microscopy work was carried out at the Canadian Centre for Electron Microscopy, a National Facility supported by McMaster University, the Canada Foundation for Innovation and NSERC. N.G. acknowledges H. Idrissi for useful discussions.

\bibliography{defectYBCO}

\end{document}